\documentclass[aps,twocolumn,prb,eqsecnum,showpacs]{revtex4}
\usepackage[dvipdfm]{graphicx}
\begin{document}
\draft
\preprint{15 August 2005}
\title{Nonlinear Lattice Relaxation of Photoexcited Diplatinum-Halide
       Chain Compounds}
\author{Jun Ohara and Shoji Yamamoto}
\address{Division of Physics, Hokkaido University,
         Sapporo 060-0810, Japan}
\date{15 August 2005}
\begin{abstract}
In order to reveal the relaxation mechanism of photogenerated
charge-transfer excitations in quasi-one-dimensional halogen-bridged
diplatinum complexes, we calculate the low-lying adiabatic potential
energy surfaces of a one-dimensional extended Peierls-Hubbard model.
High-energy excitations above the electron-hole continuum may relax into
polarons, while excitons pumped within the optical gap are self-localized
and then either decay by luminescence or divide into solitons.
Neutral solitons, charged solitons, and polarons may be simultaneously
photogenerated in a diplatinum-halide chain, which has never been observed
in any conventional platinum-halide chain.
Optical conductivity is also simulated along the decay paths for
experimental verification.
\end{abstract}
\maketitle

\section{Introduction}

   Halogen ($X$)-bridged transition-metal ($M$) chain compounds,
abbreviated as $M\!X$ chains, have been attracting much interest for
several decades.
Substituting constituent metals, halogens, ligand molecules, and counter
ions, we can widely tune their electronic structure and systematically
study low-dimensional quantum phenomena in consequence of competing
electron-electron (e-e) and electron-lattice (e-l) interactions.
\cite{G6408,W6435}
Pt$X$ chains, including Wolffram's red salt
[Pt(C$_2$H$_7$N)$_4$Cl]Cl$_2\cdot 2$H$_2$O, \cite{R2010}
exhibit a Peierls-distorted mixed-valent ground state,
whereas Ni$X$ chains, which are strongly correlated Mott insulators, have
a monovalent regular-chain structure. \cite{O2023}
Various ground \cite{B13228,R3498,Y422} and defect
\cite{B339,M5758,M5593,B5282,H16148} states were theoretically predicted
and indeed observed experimentally.
\cite{K2122,K1789,D49,H5706,O2248,K4245,K2510,O861}
The photoinduced midgap absorption of
[Pt(C$_2$H$_8$N$_2$)$_2$Cl](ClO$_4$)$_2$ caused a vigorous argument on
self-trapped nonlinear excitations, \cite{G10566,S1605,I1380,B6065}
while the gigantic optical nonlinearity of
[Ni(C$_6$H$_{14}$N$_2$)$_2$Br]Br$_2$ opened up a new way to optical
devices. \cite{K929}

   The thus-fascinating $M\!X$ family compounds have been gaining renewed
interest in recent years due to their binuclear metal analogs,
\cite{C4604,B444} abbreviated as $M\!M\!X$ chains, which exhibit quantum,
\cite{M101,Y125124,K435} thermal, \cite{K10068,I115110,Y1198}
and pressure-induced \cite{S1405,M046401,I2149,Y140102} transitions
between a wider variety of mixed-valent states. \cite{Y183,K533}
The direct $M(d_{z^2})$-$M(d_{z^2})$ overlap effectively reduces the
on-site Coulomb repulsion and therefore makes electrons more itinerant.
$M\!M\!X$ chains are indeed much more conductive than $M\!X$ chains.
\cite{K10068}
Then charge- and/or spin-carrying local excitations such as solitons
\cite{Y189} and polarons \cite{Y165113} are more and more interesting.
Thermally excited spin solitons have already been observed in
Pt$_2$(C$_5$H$_{11}$CS$_2$)$_4$I. \cite{T2169}
However, photogenerated $M\!M\!X$ defect states have neither been measured
nor been calculated yet.
Effects of metal binucleation on the relaxation of photogenerated
charge-transfer excitations are not only scientifically interesting
in themselves but must be also the key to optical switching.
Thus motivated, we simulate photoexcitation and nonlinear lattice
relaxation of diplatinum-halide chains and stimulate further
experimental explorations.

\section{Calculational Procedure}

\begin{figure}[b]
\centering
\includegraphics[width=80mm]{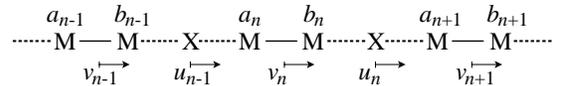}
\vspace*{-2mm}
\caption{Schematic representation of $M\!M\!X$ chains.}
\label{F:H}
\end{figure}

   We describe $M\!M\!X$ chains by the one-dimensional
$\frac{3}{4}$-filled single-band Peierls-Hubbard adiabatic Hamiltonian
\begin{eqnarray}
   &&
   {\cal H}
   =-\sum_{n,s}
     \bigl[t_{M\!X\!M}-\alpha(l_{n+1:-}+l_{n:+})\bigr]
     \bigl(a_{n+1,s}^\dagger b_{n,s}
   \nonumber \\
   &&\qquad
    +b_{n,s}^\dagger a_{n+1,s}\bigr)
    -t_{M\!M}\sum_{n,s}
     \bigl(b_{n,s}^\dagger a_{n,s}+a_{n,s}^\dagger b_{n,s}\bigr)
   \qquad
   \nonumber \\
   &&\qquad
    -\beta\sum_{n,s}
     (l_{n:-}n_{n,s}+l_{n:+}m_{n,s})
    +\frac{K_{M\!X}}{2}\sum_{n}
     \bigl(l_{n:-}^2
   \nonumber \\
   &&\qquad
    +l_{n:+}^2\bigr)
    +U_{M}\sum_{n}
     (n_{n,\uparrow}n_{n,\downarrow}+m_{n,\uparrow}m_{n,\downarrow})
   \nonumber \\
   &&\qquad
    +\sum_{n,s,s'}
     (V_{M\!M}n_{n,s}m_{n,s'}+V_{M\!X\!M}n_{n+1,s}m_{n,s'}),
   \label{E:H}
\end{eqnarray}
where
$n_{n,s}=a_{n,s}^\dagger a_{n,s}$ and
$m_{n,s}=b_{n,s}^\dagger b_{n,s}$ with
$a_{n,s}^\dagger$ and $b_{n,s}^\dagger$ creating an electron with spin
$s$ on the $M\,d_{z^2}$ orbitals in the $n$th $M\!M\!X$ unit.
$t_{M\!M}$ and $t_{M\!X\!M}$ give the intradimer and interdimer electron
hoppings, respectively.
$\alpha$ and $\beta$ describe the Peierls- and Holstein-type e-l
couplings, respectively, with $K_{M\!X}$ being the metal-halogen spring
constant.
$l_{n:-}=v_n-u_{n-1}$ and $l_{n:+}=u_n-v_n$ with $u_n$ and $v_n$ being,
respectively, the chain-direction displacements of the halogen and metal
dimer in the $n$th $M\!M\!X$ unit from their equilibrium positions.
Every diplatinum moiety, with its surrounding ligands, is not deformed.
The notation is further explained in Fig. \ref{F:H}.
We take $t_{M\!M}$ as twice $t_{M\!X\!M}$ setting $t_{M\!X\!M}$ and
$K_{M\!X}$ both equal to unity.
Since the parameter sets $(\alpha,\beta,K_{M\!X})$ and
$(a\alpha,a\beta,a^2K_{M\!X})$ are equivalent to each other with an
arbitrary constant $a$, the specific value of $K_{M\!X}$ is insignificant.
\cite{K2163}
The number of $M\!M\!X$ units, denoted by $N$, is set equal to $300$.
\vspace*{-3mm}
\begin{figure*}
\centering
\includegraphics[width=160mm]{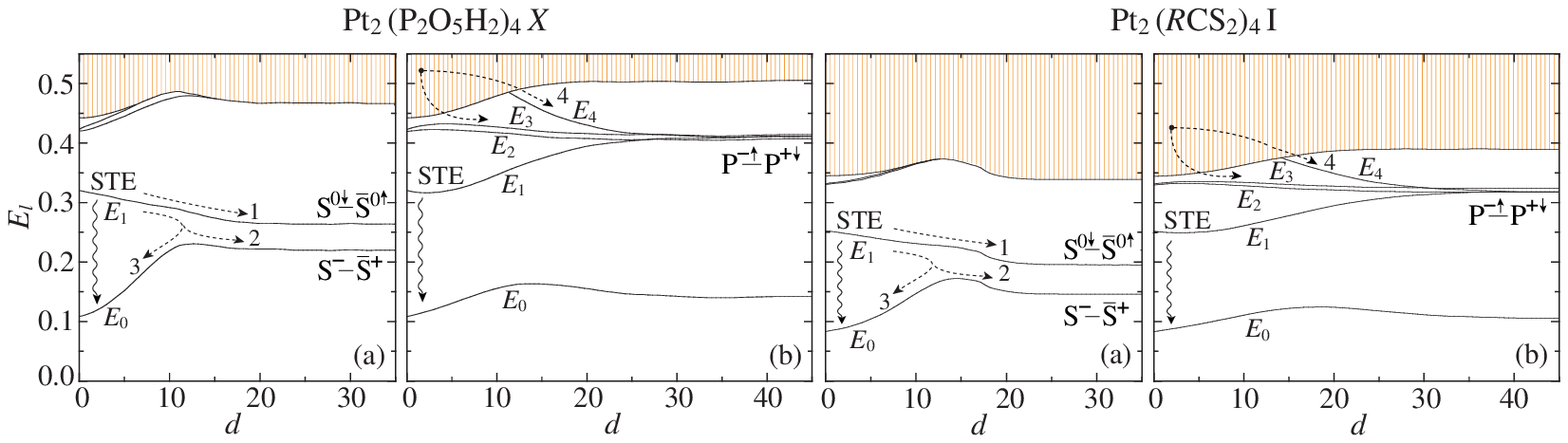}
\vspace*{-4mm}
\caption{(Color online)
         Adiabatic potential energies as functions of $d$ measured from
         the ground-state energy.
         (a): Relaxation channels connected to soliton (S)-antisoliton
         ($\bar{\mbox{S}}$) pairs;
         (b): Relaxation channels connected to polaron (P) pairs.
         Dotted (wavy) arrows suggest possible relaxation paths
         (luminescence).
         $\alpha=0.0,\beta=1.2$ and $\alpha=0.3,\beta=0.8$ for the
         P$_2$O$_5$H$_2$- and $R$CS$_2$-ligand complexes, respectively;
         $U_M=0.5$, $V_{M\!M}=0.25$, and $V_{M\!X\!M}=0.15$ in common.}
\label{F:ElWC}
\vspace*{-2mm}
\end{figure*}
\begin{figure*}
\centering
\includegraphics[width=160mm]{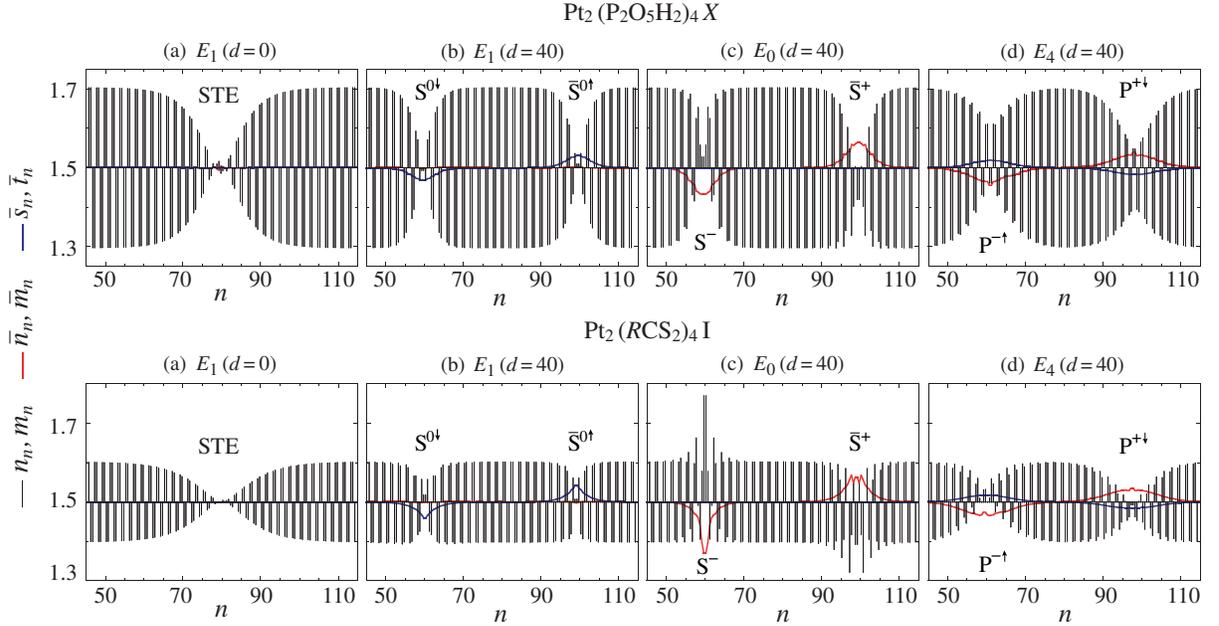}
\vspace*{-4mm}
\caption{(Color online)
         Spatial configurations of various photogenerated excitations.
         We first calculate the local electron
         ($\langle l|a_{n,\uparrow  }^\dagger a_{n,\uparrow  }|l\rangle
          +\langle l|a_{n,\downarrow}^\dagger a_{n,\downarrow}|l\rangle
           \equiv n_n,
           \langle l|b_{n,\uparrow  }^\dagger b_{n,\uparrow  }|l\rangle
          +\langle l|b_{n,\downarrow}^\dagger b_{n,\downarrow}|l\rangle
           \equiv m_n$)
         and spin
         ($\langle l|a_{n,\uparrow  }^\dagger a_{n,\uparrow  }|l\rangle
          -\langle l|a_{n,\downarrow}^\dagger a_{n,\downarrow}|l\rangle
           \equiv 2s_n,
           \langle l|b_{n,\uparrow  }^\dagger b_{n,\uparrow  }|l\rangle
          -\langle l|b_{n,\downarrow}^\dagger b_{n,\downarrow}|l\rangle
           \equiv 2t_n$)
         densities in the $l$th excited state and then extract their
         nonalternating components
         ($n_{n-1}+2n_n+n_{n+1}\equiv -4\bar{n}_n,
           m_{n-1}+2m_n+m_{n+1}\equiv -4\bar{m}_n;
           s_{n-1}+2s_n+s_{n+1}\equiv  4\bar{s}_n;
           t_{n-1}+2t_n+t_{n+1}\equiv  4\bar{t}_n$).
         In the above, the average electron density is added to the
         nonalternating components.}
\label{F:config}
\vspace*{-4mm}
\end{figure*}
\begin{figure*}
\centering
\includegraphics[width=160mm]{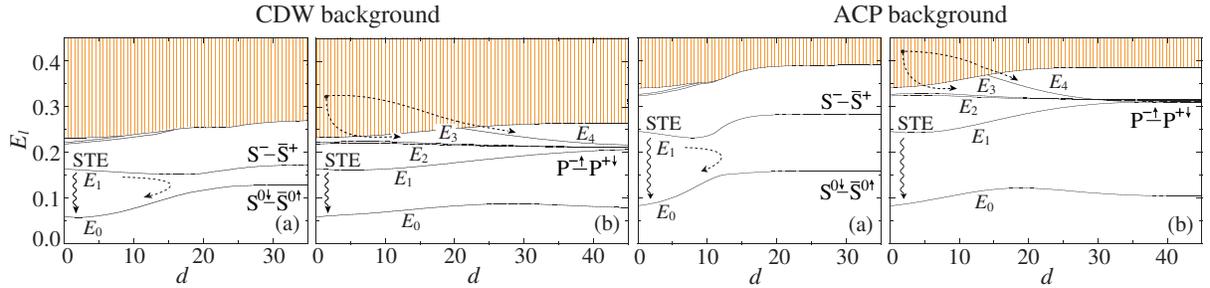}
\vspace*{-4mm}
\caption{(Color online)
         Adiabatic potential energies as functions of $d$ measured from
         the ground-state energy.
         (a): Relaxation channels connected to soliton (S)-antisoliton
         ($\bar{\mbox{S}}$) pairs;
         (b): Relaxation channels connected to polaron (P) pairs.
         Dotted (wavy) arrows suggest possible relaxation paths
         (luminescence).
         $\alpha=0.0,\beta=1.2$ and $\alpha=0.3,\beta=0.8$ for the
         CDW and ACP backgrounds, respectively;
         $U_M=1.0$, $V_{M\!M}=0.5$, and $V_{M\!X\!M}=0.3$ in common.}
\label{F:ElIC}
\end{figure*}
\begin{figure}
\centering
\includegraphics[width=80mm]{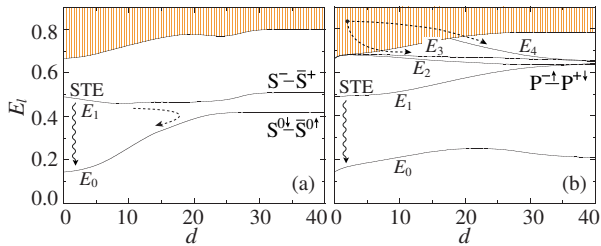}
\vspace*{-4mm}
\caption{(Color online)
         Adiabatic potential energies as functions of $d$ measured from
         the ground-state energy in $M\!X$ chains, where we employ the
         same type of Hamiltonian as Eq. (\ref{E:H}) with
         $\alpha=0.0$, $\beta=0.7$, $U_M=1.0$, and $V_{M\!X\!M}=0.3$,
         which is relevant to [Pt(C$_2$H$_8$N$_2$)$_2X$](ClO$_4$)$_2$
         ($X=\mbox{Cl},\mbox{Br}$).
         (a): Relaxation channels connected to soliton (S)-antisoliton
         ($\bar{\mbox{S}}$) pairs;
         (b): Relaxation channels connected to polaron (P) pairs.
         Dotted (wavy) arrows suggest possible relaxation paths
         (luminescence).}
\label{F:ElMX}
\end{figure}

   The existent diplatinum-halide chain compounds are classified into two
groups:
 $A_4$[Pt$_2$(P$_2$O$_5$H$_2$)$_4X$]$\cdot$$n$H$_2$O
($X=\mbox{Cl},\mbox{Br},\mbox{I}$; $A=\mbox{Li},\mbox{Cs},\cdots$)
\cite{C4604,C409} and
 Pt$_2$($R$CS$_2$)$_4$I
($R=\mbox{C}_n\mbox{H}_{2n+1}$). \cite{B444,I387}
The former structurally resembles the $M\!X$ conventional and exhibits a
ground state with halogen-sublattice dimerization:
$-X^{-}\!\cdots\mbox{Pt}^{2+}\mbox{Pt}^{2+}\!\cdots
 X^{-}\!-\mbox{Pt}^{3+}\mbox{Pt}^{3+}\!-X^{-}\!\cdots$,
which is referred to as the charge-density-wave (CDW) state, where the
intrasite e-l coupling $\beta$ is dominant.
The latter possesses a distinct ground state with metal-sublattice
dimerization,
$\cdots\mbox{I}^{-}\!\cdots\mbox{Pt}^{2+}\mbox{Pt}^{3+}\!-\mbox{I}^{-}
 \!-\!\mbox{Pt}^{3+}\mbox{Pt}^{2+}\!\cdots\mbox{I}^{-}\!\cdots$,
which is referred to as the alternate charge-polarization (ACP) state,
where the intersite e-l coupling $\alpha$ is significant.
Platinum-halide chains exhibit intermediate e-e interactions
($V_{M\!X\!M}\ll U_M\alt t_{M\!X\!M}$) \cite{W6435,M5758} and platinum
binucleation should reduce the on-site repulsion.
Thus the Pt$_2X$ Coulomb parameters are, unless otherwise noted, set for
$U_M=0.5$, $V_{M\!M}=0.25$, and $V_{M\!X\!M}=0.15$.
The e-l coupling constants are taken in two ways as
$\alpha=0.0,\beta=1.2$ and $\alpha=0.3,\beta=0.8$, which are relevant to
the P$_2$O$_5$H$_2$- and $R$CS$_2$-ligand complexes and indeed give
the CDW and ACP ground states, respectively, under the above Coulomb
parametrization. \cite{Y13}

   Photogenerated charge-transfer excitations, spreading over the chain
at first, are self-localized into excitons, solitons, and polarons.
\cite{O2023}
The whole relaxation scenario is describable with a trial wave function
\cite{M5758,Y165113,I1088}
\begin{equation}
   l_{n:\pm}
  =(-1)^n l_0
   \Biggl[
    1+\delta l
      \Bigl({\rm tanh}\frac{|n\pm\delta n|-d/2}{\sigma_n\xi}-1\Bigr)
   \Biggr],
   \label{E:WF}
\end{equation}
where $l_0$ is set equal to the halogen-ion displacement in the uniformly
distorted CDW ground state, while the rest of the variational parameters
are determined so as to minimize the energy of the lowest-lying excited
state.
Once the CDW state is photoexcited into the Frank-Condon state, which
still sits at $\delta l=0$, the uniform bond alternation begins to be
locally deformed.
Increasing $\delta l$ with $d$ fixed to zero represents the
self-localization of a charge-transfer exciton.
The self-trapped exciton (STE) may further divide into a pair of local
defects with increasing $d$.
There is also a possibility of a higher-energy pumped electron-hole (e-h)
pair directly splitting into distant defects in a pair, with
simultaneously increasing $\delta l$ and $d$.
$\xi$ corresponds to the extent of a local defect, whereas $\delta n$
allows the neighboring metal sites to behave unequally around the defect
center.
$\sigma_n$ takes $\pm$ according to relaxation channels.
It is simply set equal to unity for solitonic defects, while it is defined
as $\sigma_n=\mbox{sgn}(|n\pm\delta n|-d/2)$ for polaronic defects.
In both cases, the key variable $d$ indicates the interdefect distance.
We optimize $\delta l$, $\xi$, and $\delta n$ at each $d$.

   With given $l_{n:\pm}$, we solve the Hartree-Fock (HF) Hamiltonian
${\cal H}_{\rm HF}$ and obtain the low-lying states
$|l\rangle^{\quad}_{\rm HF}$ of energy $E_l^{\rm HF}$ ($l=0,1,2,\cdots$).
Then we consider refining the description of excited states, which may be
generally given as
\begin{equation}
   |l\rangle
   =\sum_s
    \sum_{\epsilon_{\mu,s}\leq\epsilon_{\rm F}}
    \sum_{\epsilon_{\nu,s}>\epsilon_{\rm F}}
    f(\mu,\nu, s;l)c_{\nu,s}^\dagger c_{\mu,s}|0\rangle^{\quad}_{\rm HF},
\end{equation}
where $\epsilon_{\rm F}$ is the Fermi energy and $c_{\lambda,s}^\dagger$
creates an electron with spin $s$ in the $\lambda$th HF eigenstate of
energy $\epsilon_{\lambda,s}$.
The HF scheme describes any excited state as a single Slater determinant,
that is, $f(\mu,\nu,s;l)=\delta_{\mu\nu s,l}$, taking no account of the
residual interaction ${\cal H}-{\cal H}_{\rm HF}\equiv{\cal V}$.
Full diagonalization of ${\cal H}$ on the basis of $|l\rangle$ must be
the best way to deal with the excitonic effect.
However, even at $N=120$, for instance, such a calculation costs eight
gigabyte memory and a hundred hours, which means spending more than a year
in optimizing the wave function (\ref{E:WF}) at every fixed $d$.
In an attempt to take the excitonic effect into calculation more
efficiently, we may consider neglecting all the off-diagonal elements of
${\cal V}$ and correcting the energy scheme perturbationally. \cite{M5758}
Then the $l$th excited-state energy is expressed as
$E_l=E_0+\epsilon_{\nu,s}-\epsilon_{\mu,s}
+\!\!\!\!\!\!\quad^{\quad}_{\rm HF\!}\langle l|
 {\cal V}|l\rangle^{\quad}_{\rm HF}$,
where $E_0\equiv E_0^{\rm HF}$.
Such a perturbational treatment of the excitonic effect is not only
practical but also fairly quantitative under the not-so-strong electronic
correlation of our present interest.
We have indeed confirmed for short chains that the thus-obtained energies
$E_l$ well approximate those of excited states of the
configuration-interaction (CI) type (see Appendix \ref{A:EE}).
This is not the case with strongly correlated nickel complexes.
\cite{O1571}
The following variational calculation is carried out so as to minimize
$E_1$ with respect to $\delta l$, $\xi$, and $\delta n$ at each $d$.
When we take particular interest in the $l$th energy surface, $E_l$ may be
minimized instead.
However, the whole energy scheme is not sensitive to the variational
target and remains unchanged visually.

\section{Adiabatic Potential Energy Surfaces}

   Figure \ref{F:ElWC} presents the thus-calculated energy surfaces,
the left two of which are relevant to
$A_4$[Pt$_2$(P$_2$O$_5$H$_2$)$_4X$]$\cdot$$n$H$_2$O, while the right two
of which are to Pt$_2$($R$CS$_2$)$_4$I.
The electron-hole continuum of Pt$_2$($R$CS$_2$)$_4$I lies lower in energy
than that of $A_4$[Pt$_2$(P$_2$O$_5$H$_2$)$_4X$]$\cdot$$n$H$_2$O.
In fact Pt$_2$(CH$_3$CS$_2$)$_4$I exhibits much smaller optical gap than
conventional $M\!X$ chain compounds and even shows metallic conduction at
room temperature. \cite{K10068}
The relaxation channels of the P$_2$O$_5$H$_2$- and $R$CS$_2$-ligand
complexes are qualitatively the same.
Fully trapped excitons [Fig. \ref{F:config}(a)], unless decay by
luminescence, are dissociated into soliton (S)-antisoliton
($\bar{\mbox{S}}$) pairs [Figs. \ref{F:config}(b) and \ref{F:config}(c)],
whereas there is no possibility of their relaxing into polaron (P) pairs.
$\mbox{P}^{-\uparrow}\!-\mbox{P}^{+\downarrow}$ pairs
[Fig. \ref{F:config}(d)] may be created from higher-energy excited states.
No energy barrier between an STE and any $\mbox{S}-\bar{\mbox{S}}$ pair, 
which is not the case with Pt$X$ chains, \cite{M5758,I1088} should result
in very short decay time of STEs and long life time of
$\mbox{S}-\bar{\mbox{S}}$ pairs.
The soliton and antisoliton in any $\mbox{S}-\bar{\mbox{S}}$ pair have
localized wave functions and their overlap rapidly decreases with
increasing $d$.
Therefore, instantaneous charge transport between far distant S and
$\bar{\mbox{S}}$ is hardly probable and tunneling between the energy
surfaces $E_1$ and $E_0$ is restricted to the region of moderately small
$d$.

   Under the relatively weak correlation relevant to Pt$_2X$ chains,
charged soliton pairs $\mbox{S}^-\!-\bar{\mbox{S}}^+$ are lower in energy
than neutral soliton pairs
$\mbox{S}^{0\downarrow}\!-\bar{\mbox{S}}^{0\uparrow}$. \cite{O115112}
Figures \ref{F:config}(b) and \ref{F:config}(c) show that solitons and
antisolitons on the energy surface $E_1$ carry net spins, whereas those on
the energy surface $E_0$ convey net charges.
As the Coulomb repulsion grows, the polaronic channel qualitatively
remains unchanged, while the solitonic one significantly varies.
Figure \ref{F:ElIC} indeed shows that $\mbox{S}^-\!-\bar{\mbox{S}}^+$
pairs are possibly higher in energy than STEs as well as
$\mbox{S}^{0\downarrow}\!-\bar{\mbox{S}}^{0\uparrow}$ pairs.
Increasing on-site Coulomb repulsion reduces the Peierls gap, especially
on the CDW background.
Figure \ref{F:ElIC} is reminiscent of the situation in Pt$X$ chains,
\cite{M5758,I1088} which is reproduced in Fig. \ref{F:ElMX}.
Photoinduced midgap absorption of
[Pt(C$_2$H$_8$N$_2$)$_2X$](ClO$_4$)$_2$ ($X=\mbox{Cl},\mbox{Br}$)
\cite{K4245,O861} was actually attributed to neutral solitons.
\cite{O2023,I1088}
In comparison with PtCl and PtBr chains, PtI chains have a larger
supertransfer energy $t_{M\!X\!M}$ (Ref. \onlinecite{G6408}) and thus
exhibit effectively suppressed e-e interactions.
Consequently there appear charged solitons instead of neutral ones in
photoexcited [Pt(C$_2$H$_8$N$_2$)$_2$I](ClO$_4$)$_2$. \cite{O2248}
Pt$X$ compounds generally exhibit further absorption bands due to
polarons within the gap, \cite{O2023,D13185} provided the excitation
energy is higher than the charge-transfer gap $\mit\Delta$.
However, there is no signal of neutral and charged solitons being
simultaneously photogenerated in Pt$X$ chains, probably because either
$\mbox{S}^{0\downarrow}\!-\bar{\mbox{S}}^{0\uparrow}$ or
$\mbox{S}^-\!-\bar{\mbox{S}}^+$ pairs are necessarily higher in energy
than STEs. \cite{I1088}
Figure \ref{F:ElWC} promises dramatic observations:
{\it Neutral solitons, charged solitons, and polarons may be
simultaneously detected in photoexcited diplatinum-halide chain
compounds.}
The decay time of luminescent STEs, $\tau$, is a few hundred picoseconds
in [Pt(C$_2$H$_8$N$_2$)$_2$Cl](ClO$_4$)$_2$, \cite{T12716,W8276} which is
much shorter than the radiative lifetime and therefore comes from the
dissociation to $\mbox{S}^{0\downarrow}\!-\bar{\mbox{S}}^{0\uparrow}$
pairs.
In $A_4$[Pt$_2$(P$_2$O$_5$H$_2$)$_4X$]$\cdot$$n$H$_2$O,
the $\mbox{S}^{0\downarrow}\!-\bar{\mbox{S}}^{0\uparrow}$ path is likely
open to an STE without any potential barrier in between and therefore STEs
must decay still faster possibly with $\tau\ll 100\,\mbox{ps}$.
\begin{figure*}
\centering
\includegraphics[width=160mm,bb=0 0 766 364]{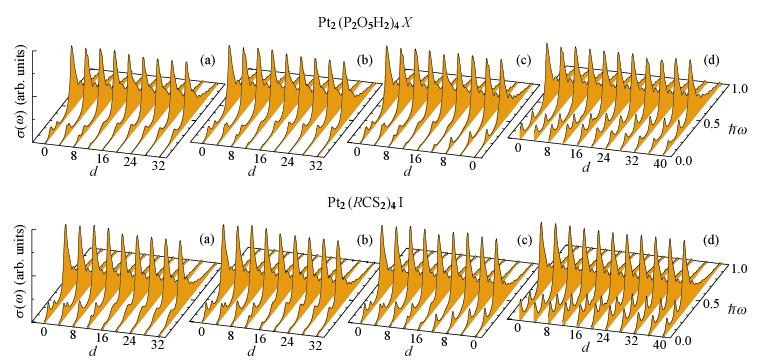}
\vspace*{-3mm}
\caption{(Color online)
         Photoinduced absorption spectra along possible relaxation paths:
         (a) $\mbox{STE}\rightarrow
              \mbox{S}^{0\downarrow}\!-\bar{\mbox{S}}^{0\uparrow}$;
         (b) $\mbox{STE}\rightarrow
              \mbox{S}^-\!-\bar{\mbox{S}}^+$;
         (c) $\mbox{STE}\rightarrow\mbox{S}-\bar{\mbox{S}}
              \rightarrow\mbox{CDW}$;
         (d) $\mbox{e}-\mbox{h}\rightarrow
              \mbox{P}^{-\uparrow}\!-\mbox{P}^{+\downarrow}$.}
\label{F:OC}
\end{figure*}
\begin{figure*}
\centering
\includegraphics[width=160mm,bb=0 0 474 566]{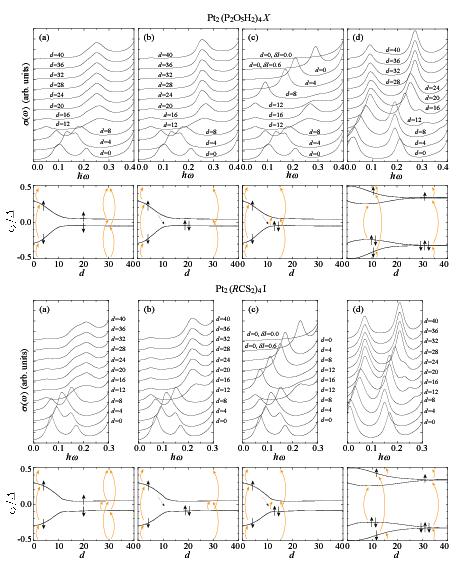}
\vspace*{-4mm}
\caption{(Color online)
         The low-energy part of Fig. \ref{F:OC} is scaled up, together
         with the intragap level scheme and occupancy, where round arrows
         denote optically allowed excitations, while dotted arrows suggest
         tunneling between different energy surfaces.}
\label{F:OCillust}
\vspace*{-4mm}
\end{figure*}

\section{Photoinduced Absorption Spectra}

   In order to encourage time-resolved optical measurements on Pt$_2X$
chains, we calculate photoinduced absorption spectra as functions of the
interdefect distance $d$.
The real part of the optical conductivity on the potential energy surface
$E_l$ is represented as
\begin{equation}
   \sigma(\omega)
    =\frac{\pi}{N\omega}\sum_{l'}
   \bigl|\langle l'|{\cal J}|l\rangle\bigr|^2
   \delta(\widetilde{E}_{l'}-E_l-\hbar\omega),
\end{equation}
where the current operator ${\cal J}\equiv\sum_{n,s}j_{n,s}$ is given by
\begin{eqnarray}
   &&
   j_{n,s}
   =\frac{{\rm i}e}{\hbar}c_{M\!X\!M}
    \bigl[t_{M\!X\!M}
     -\alpha(l_{n+1:-}+l_{n:+})\bigr]
     \bigl(a_{n+1,s}^\dagger b_{n,s}
   \ \nonumber \\
   &&\ 
   -b_{n,s}^\dagger a_{n+1,s}\bigr)
   +\frac{{\rm i}e}{\hbar}
    c_{M\!M}t_{M\!M}
    \bigl(b_{n,s}^\dagger a_{n,s}-a_{n,s}^\dagger b_{n,s}\bigr),
\end{eqnarray}
with $c_{M\!M}$ and $c_{M\!X\!M}$ being the average $M$-$M$ and
$M$-$X$-$M$ distances, respectively, and set for $c_{M\!X\!M}=2c_{M\!M}$.
When we take $|l\rangle_{\rm HF}^{\quad}$ for $|l\rangle$ with the
perturbationally corrected energy scheme $E_l$, we obtain
$|l'\rangle=c_{\nu',s'}^\dagger c_{\mu',s'}|l\rangle_{\rm HF}^{\quad}$ and
$\widetilde{E}_{l'}=E_l^{\rm HF}+\epsilon_{\nu',s'}-\epsilon_{\mu',s'}
+\langle l'|{\cal V}|l'\rangle$.
Figures \ref{F:OC}(a)$-$\ref{F:OC}(d) present the optical conductivity
spectra along the relaxation paths labeled 1$-$4 in Fig. \ref{F:ElWC},
where (a)$-$(c) are characteristic of solitonic excitations via STEs,
while (d) of a polaronic one from the e-h continuum.
The intragap absorption bands visualize various nonlinear lattice
relaxation paths distinguishably, which are scaled up and interpreted in
Fig. \ref{F:OCillust}.

   In Fig. \ref{F:OCillust}(a), an STE splits into far distant
$\mbox{S}^{0\downarrow}$ and $\bar{\mbox{S}}^{0\uparrow}$,
while in Fig. \ref{F:OCillust}(b), an STE relaxes into far distant
$\mbox{S}^-$ and $\bar{\mbox{S}}^+$ via quantum tunneling.
In Pt$X$ chains, $\mbox{S}^-\!-\bar{\mbox{S}}^+$ pairs, instead of
$\mbox{S}^{0\downarrow}\!-\bar{\mbox{S}}^{0\uparrow}$ pairs, lie in the
same potential energy surface as STEs', but they can not be reached due to
their high energy.
Here in Pt$_2X$ chains, STEs can nonradiatively decay into both neutral-
and charged-soliton pairs and therefore the early decrease of their
luminescence intensity should look double-exponential.
The optical conductivity spectra are informative about the soliton charge.
With increasing $d$, the exciton-to-band infrared absorption peak moves
upward, whereas the intra-exciton midgap one downward.
After passing each other, the former merges into the soliton-to-band
absorption, while the latter fades out with $\mbox{S}$ and
$\bar{\mbox{S}}$ going away from and less overlapping with each other.
Such a scenario is common to both
$\mbox{S}^{0\downarrow}\!-\bar{\mbox{S}}^{0\uparrow}$ and
$\mbox{S}^-\!-\bar{\mbox{S}}^+$ paths, but their long-time spectra look
different.
The neutral-soliton-to-band absorption spectrum consists of two close
peaks and is thus broader than the charged-soliton-to-band one.
For Pt$_2$($R$CS$_2$)$_4$I, it is remarkably humped due to the two
intragap soliton levels lying relatively far apart from each other.
On the ACP background, the intragap soliton levels are more sensitive to
the e-e interactions. \cite{O250}
Pt$_2$($R$CS$_2$)$_4$I exhibits another surviving absorption in the
infrared region, which is attributable to the electron transfer between
S and $\bar{\mbox{S}}$.
This is because of larger $\xi$ in Pt$_2$($R$CS$_2$)$_4$I
(see Fig. \ref{F:config}).
Solitons are more delocalized on the ACP background than on the CDW
background.
Figure \ref{F:OCillust}(c) describes a nonradiative geminate
recombination through the solitonic state, where the intra-exciton
absorption peak turns upward and merges with the background interband
absorption.

   Figure \ref{F:OCillust}(d) depicts polaronic excitations, which are
efficiently generated through high-energy pumping.
There are four potential energy surfaces leading to
$\mbox{P}^{-\uparrow}\!-\mbox{P}^{+\downarrow}$ pairs.
Although STEs lie in one of them, they have no chance of relaxing into
polarons.
When charge-transfer excitations split into polaron pairs, the intragap
absorption spectra at small $d$, that is, the early observations,
significantly vary with their relaxation paths, but they all converge to
well-separated two bands.
The intra-polaron absorption is higher in energy and stronger in intensity
than the polaron-to-band one.

\section{Summary}

   We have calculated the nonlinear lattice relaxation paths of
photogenerated charge-transfer excitations in diplatinum-halide chain
compounds and ``observed" their optical conductivity spectra.
There is a possibility of
{\it neutral solitons, charged solitons, and polarons coexisting in
photoexcited Pt$_2X$ chains} especially of the $R$CS$_2$-ligand type,
which is never the case with conventional Pt$X$ chains.
The optical-conductivity spectra due to polaron pairs generally comprise
well-separated two bands, while those due to soliton pairs mainly consist
of a single band but its structure is rather varied with the ground state
and the relaxation path.
The $\mbox{S}^-$-to-band absorption is truly single-peaked, while the
$\mbox{S}^0$-to-band absorption gives a broader band with a shoulder in
the lower-energy side.
In Pt$_2$($R$CS$_2$)$_4$I, the S-to-$\bar{\mbox{S}}$ absorption is long
surviving and detectable in the infrared region in addition.

   Indeed there are many similarities between the Pt$_2X$ and Pt$X$
relaxation mechanisms, but their solitonic channels qualitatively differ.
{\it In Pt$_2X$ chains, both neutral- and charged-soliton paths are open
to photogenerated charge-transfer excitons} and soliton pairs stably
survive longer life time.
The present study is motivated in part by the pioneering calculations on
Pt$X$ chains, \cite{M5758} where no observable to be measured was
presented, however.
Nowadays, the time-resolved optical spectroscopy technique has been much
more refined than before and the full spectrum can be analyzed with
femtosecond precision. \cite{I057401}
Such circumstances have stimulated us to have the idea of calculating
Fig. \ref{F:OC}, spending a week on each run along the relaxation paths.
Even the photoinduced midgap absorption of a prototypical platinum-halide
chain compound [Pt(C$_2$H$_8$N$_2$)$_2$Cl](ClO$_4$)$_2$ \cite{K1789,S3066}
is not yet completely solved in spite of the enthusiastic argument.
\cite{D49,K4245,S1605,I1380,I1088}
Diplatinum-halide chain compounds comprise two groups with distinct
mixed-valent ground states and their electronic states are much more
tunable.
Comparative measurements, including photoinduced absorption, luminescence,
and electron spin resonance, on the
P$_2$O$_5$H$_2$- and $R$CS$_2$-ligand $M\!M\!X$ complexes must reveal
their intrinsic nonlinear lattice relaxation mechanism and may even give a
key to unsettled issues on $M\!X$ complexes.

\acknowledgments

   The authors thank K. Iwano for fruitful discussions.
This work was supported by the Ministry of Education, Culture, Sports,
Science, and Technology of Japan.

\begin{appendix}
\section{Comparison between the HF and CI Schemes}
\label{A:EE}

\begin{figure}
\centering
\includegraphics[width=80mm]{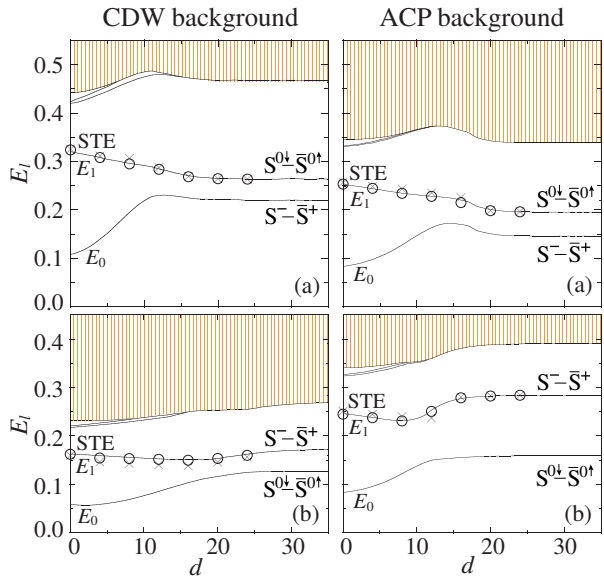}
\vspace*{-4mm}
\caption{(Color online)
         Relaxation paths leading to soliton (S)-antisoliton
         ($\bar{\mbox{S}}$) pairs.
         $\alpha=0.0,\beta=1.2$ and $\alpha=0.3,\beta=0.8$ for the
         CDW and ACP backgrounds, respectively.
         (a) $U_M=0.5$, $V_{M\!M}=0.25$, $V_{M\!X\!M}=0.15$;
         (b) $U_M=1.0$, $V_{M\!M}=0.5$, $V_{M\!X\!M}=0.3$.}
\label{F:EE}
\end{figure}

   Adiabatic potential energies of short chains ($N=50$) are calculated by
the CI scheme ($\times$) as well as the perturbationally refined HF method
($\circ$) and are compared with Figs. \ref{F:ElWC}(a) and \ref{F:ElIC}(a)
in Fig. \ref{F:EE}.
\end{appendix}

\end{document}